# PAVING THE WAY FOR THE SUCCESS OF MAGNETO-OPTICS

Hanns-Ulrich Habermeier    MPI- FKF Stuttgart, Germany

Abstract: Quantitative high resolution magneto-optical imaging of flux structures as developed in the period 1970-1980 gained much popularity after the discovery of high temperature superconductivity in cuprates. In this historical review the developments of the early days are described with applications of this technique to study flux penetration in type-I and the determination of pinning forces in type-II superconductors.

Key words:    flux penetration, quantitative magneto-optics, flux-line pinning

## 1. INTRODUCTION

The discovery of high temperature superconductors [HTS] generated an avalanche of interests in different fields of condensed matter physics including vortex matter physics. Especially in this field a plethora of new phenomena emerged partially based on the strong crystalline anisotropy of HTS materials, their short anisoptropic coherence lengths and the much larger thermal energies involved as compared to the classical low $T_C$ superconductors [SC]. Similarly to the investigations of the magnetic properties of classical SC's there is a demand to complement informations gained from bulk measurements including magnetization, magnetotransport and heat capacity by a microscopic picture of flux penetration, flux structures and flux dynamics. In this context the concepts developed during the 'pre-HTS period' to study flux properties locally deserve special attention. Prior to 1970 several approaches have been used to study flux structures locally, ranging from Bitter techniques either with ferromagnetic [1] or superconducting [2,3,4] particles, to neutron scattering [5] and magneto-optical imaging [MOI] based on the principle of the Faraday effect [6,7]. A historical breakthrough was the development of a high resolution Bitter technique by Träuble and Essmann [8] and subsequently by Sarma and Moon [9] where not only the existence of flux-lines could be experimentally proven but also fundamental properties of the flux-line lattice such as its symmetry, defect formation and the role of flux-line dislocations have been studied for the first time at a sub - µm scale [8].

In this historical review the fundamentals of the magneto-optical technique are described and the development during the period 1970 – 1980 is treated in some detail. During this period there have been three major developments, the introduction of the high resolution MOI by Kirchner [10], the study of time-dependent and current-induced phenomena by Huebener [11] and the quantitative determination of flux density gradients in type – II superconductors by Habermeier [12]. It is the purpose of this paper to give some examples of the early experiments using type – I SC's and focus on the development of MOI towards a quantitative tool to investigate flux-density gradients and pinning forces in type – II SC's.

## 2. FUNDAMENTALS OF MAGNETO-OPTICAL IMAGING OF MAGNETIC STRUCTURES IN SUPERCONDUCTORS

MOI of flux structures is based on the magneto-optical Faraday effect which describes empirically the rotation, $\alpha$, of the vector of linearly polarized light propagating parallel to the magnetic field direction over a distance, t, through a medium with longitudinal optical birefringence in a magnetic field, H,

$$\alpha = V(\omega)tH \quad (1)$$

The eigenmodes of light propagation [ left and right circular polarized ] have a different index of refraction $\Delta n = n_L(\omega) - n_R(\omega)$ directly proportional to the expectation value of the magnetic moment along the propagation axis. This rotation of the polarization plane in conjunction with polarization optics is used in MOI. In a typical geometry for MOI the light passes through MO layer [MOL], is reflected at the sample surface and passes a second time through the MOL, thus doubling the rotation angle. Most of the magneto-optical materials, however, have a high absorption coefficient, $\beta$, consequently, the intensity of the reflected light after passing through the analyser is

$$I = 4(I_0 V t H)^2 \exp(-2\beta t) \quad (2)$$

With a maximum for $t = 1/\beta$ [13]. To optimise the effect, the figure of merit is $V/\beta$. To determine the conditions for a MO contrast [10], the corresponding electrical field vector, F, is split into components parallel, $|F|\sin\alpha$, and perpendicular, $|F|\cos\alpha$, to the transmitting direction of the analyser; the condition for a MO contrast is therefore

$$T_\parallel |F|^2 \sin^2\alpha > T_\perp |F|^2 \cos^2\alpha \quad (3)$$

where $T_{\|}$ and $T_{\perp}$ denote the transmittance of light polarized parallel and perpendicular to the transmission direction of the analyser. For small angles, therefore, the contrast condition is

$$\alpha > (T_{\perp}/T_{\|})^{1/2} \quad (4)$$

Originally, this concept of MOI was introduced by Alers 1957 [6] using $Ce^{3+}$ salts which had a small absorption coefficient and relatively high Verdet's constants of $V \sim 5 \cdot 10^{-5}$ $^0$/mTµm. This technique allows a spatial resolution of ¼ mm due to the limitations given by the relatively thick MO layers [14] required to fulfill the contrast conditions according to equ. (4). DeSorbo [7] developed this technique further, allowing the study of dynamical phenomena, even motion pictures and high speed photography became possible. The spatial resolution, however, was still confined to ¼ mm. Irrespective of the meanwhile developed Träuble-Essman decoration with sub-µm resolution there was still a need for improvement of the MOI partially due to the flexibility of MOI, partially due to the sophistication of the decoration technique.

## 3. HIGH RESOLUTION MAGNETO-OPTICAL IMAGING

The main drawback of MOI developed so far was the low spatial resolution mainly determined by the MOL thickness. The requirement for a high resolution MOI technique, therefore, is the availability of thin (< 1µm) MO layers with a high Verdet constant. Cross-fertilization for these efforts came from a different field of solid state physics. Methfessel [15] as well as Wachter and Busch [16] studied the optical properties of the divalent Europiumchalkogenides EuO, EuS, and EuSe and found Verdet's constants of the order of $10^{-2} - 10^{-1}$ $^0$/ mTµm at 4.2K i.e. larger by two orders of magnitude compared to the $Ce^{3+}$ salts, thus allowing a reduction of the MOL thickness to less than 1µm. Unfortunately, these compounds order ferromagnetically at 66.8K and 16.9K for EuO and EuS, respectively, or have a metamagnetic transition at 4.6K to antiferromagnetism and 2.8K to ferromagnetism in the case of EuSe. Inorder to reveal the real flux structures of a SC ferromagnetism had to be suppressed in the Eu-chalkogenide films. It was the breakthrough by Kirchner [10] leading to high resolution MOI who pursuit the concept of admixing paramagnetic $EuF_2$ to EuS thus achieving two major goals: (i) suppression of ferromagnetism while maintaining a high Verdet constant and (ii) reduction of the absorption coefficient of a plain EuS film by $EuF_2$ admixture and thus enlarging V/β substantially. Systematic studies to optimize the EuS/EuF$_2$ ratio revealed an EuF$_2$ admixture necessary between 10 and 60 wt-% depending on the evaporation conditions [17]. Further optimisation principles for the contrast as developed by Kirchner are the enlargement of $\alpha$ by interference – i.e. using the MOL as antireflecting coating – and partial polarisation as described in detail in [10].

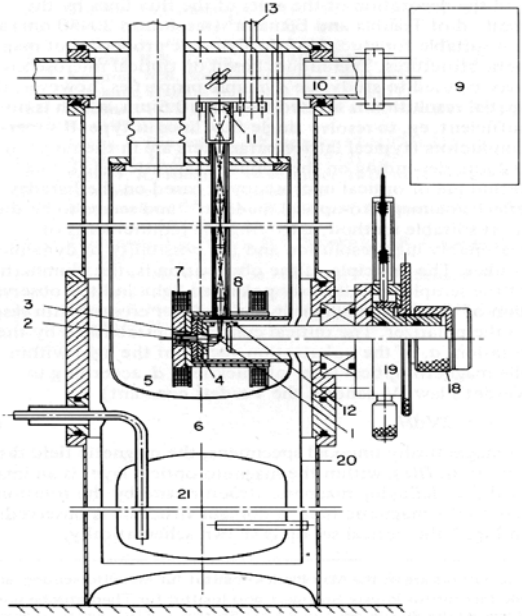

*Figure 1 Drawing of the Faradaycryostat [18]*

In addition to the development of appropriate MOL's with high Faradayrotation the second prerequisite for high resolution MOI is an imaging system allowing access to low temperatures combined with a small working distance for the objective lens of a low temperature microscope. There have been several solutions for this problem using either beam splitters to illuminate the sample and a microscope objective in the vacuum chamber or the use of a conventional room temperature microscope with a cryostat equipped with a window to separate the cold stage from the microscope. In order to achieve spatial resolutions of 1µm the first approach is preferable. The solution realized by Habermeier is given in Fig. 1 [18]. In this rather rigid construction the optical path is well aligned, polarizer and analyzer can be separately

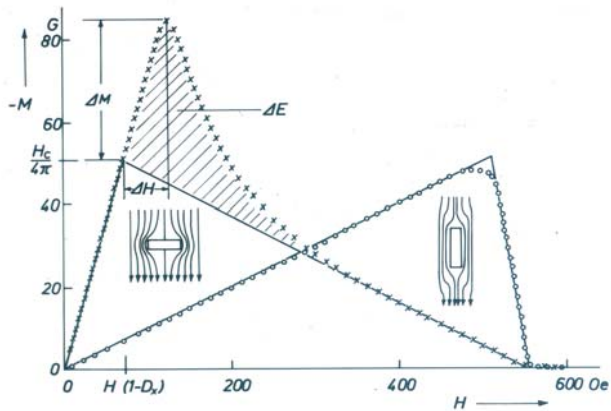

*Figure2 Magnetization curve of Pb disc at 4.2K*

adjusted and a dielectric mirror is the only optical component between polarizer and analyzer. The advantage of this set-up is the possibility to change specimens without warming-up the system and changing the relative positions of polariser and analyser.

## 4. FLUXPENETRATION IN TYPE– I SUPER-CONDUCTORS

The magnetization of a superconductor placed in a magnetic field $H_{ex}$ gives rise to a stray field $H_S$. If at the periphery the local resulting field $H_{ex} + H_S$ exceeds the entry field $H_{ent}$ normal conducting domains start to enter the SC. The entry field includes the thermodynamic critical field $H_C$ and an additional component given by a possible barrier for flux penetration. The state of coexistence of normal and superconducting regions in a type – I SC is called intermediate state and is stable in the range $(1-D) H_C < H_{ex} < H_C$. Fig. 2 shows the measured magnetization curve of a lead disc in field orientations perpendicular and parallel to the plane surfaces. The deviations from the ideal curve as described by $M = -[1/4\pi(1-D)]H_{ex}$ for $0 < H_{ex} < (1-D)H_C$ and $M = (H_{ex} - H_C)/4\pi D$ in the region $(1-D)H_C < H_{ex} < H_C$ are related to the domain structure and flux penetration. Early MOI experiments by Baird [19] suggested that in the field interval between $(1-D)H_C$ and the maximum of the curve an intermediate state at the edge of the specimen is established consisting in a fingerlike structure pointing towards the center of the sample. At the maximum macro fluxlines are expelled from these fingers and drift into the interior of the specimen where a coagulation of the domains takes place and the typical flux patterns of type-I SC's are formed. With increasing field the flux free belt decreases and disappears when the magnetization curve meets the theoretical one. In Fig. 3 this situation is represented for a Pb disc with 4 mm diameter and 1 mm thickness. Experimental studies by Habermeier [20] and Essmann et. al. [21] as well as theoretical studies by Fortini [24], Kronmüller and Riedel [22] and Fähnle [23] have shown that the SN- wall energy, their interaction with lattice defects and especially the geometry of the sample determines the shape of the magnetization curve. The irreversibility of the magnetization curve and the excess energy ΔE in Fig.2 is basically due to the dissipative movement of the macro flux-lines across the flux free belt.

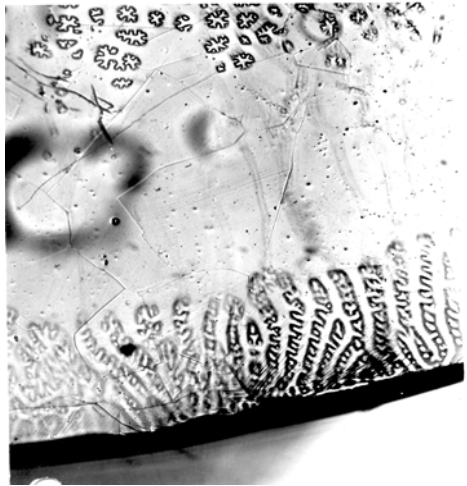

*Fig. 3 Domain structure of e Pb disc at 1.2K [ decoration technique ]*

The flux pattern of Fig. 3 changes completely if a small notch is introduced at the periphery of the sample. Flux nucleation exclusively occurs at the notch due to the enhanced inhomogeneous stray field and a large flux free belt is formed. As pointed out [25] this could be used in principle to generate flux free regions for a potential application in superconductor device technology.

## 5. MOI OF FLUX PATTERNS IN TYPE-II SUPERCON-DUCTORS

In the mixed state of type-II SC's the flux-line distance is $d=[2\Phi_0/B\sqrt{3}]^{1/2}$ with values far below the resolution of the MOI technique. Consequently the efforts to use the MOI technique were confined before 1975 to investigations of type-I SC's and some rather qualitative low resolution observations of flux structures in Nb-foils. [26]. The first application of MOI for quantitative studies emerged from the one dimensional Friedel model [27] for the determination of pinning forces $f_p$

$$f_p = - B(\partial H/\partial B)_{rev}\partial B/\partial r$$

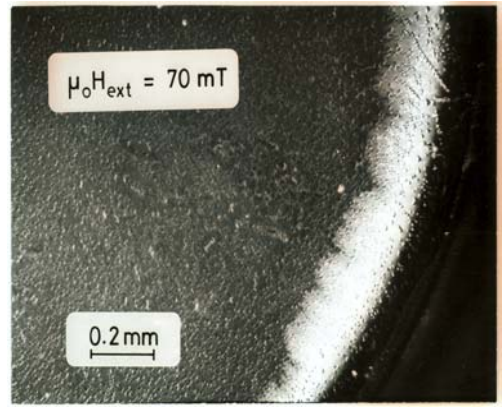

*Figure 4. Domain structure of e Nb disc at 4.5K   [ MOI ]*

If a direct method for the determination of $\partial B/\partial r$ could be established, the direct local measurement of $f_p$ would be possible. The concept using MOI as introduced by Habermeier and Kronmüller [28] is based on an averaging of local flux densities at a μm scale and the measurement of their spatial variation.

In Fig. 4 the flux penetration pattern of a zero field cooled (110) Nb-disc in the mixed state is given indication the radial flux penetration. Fig. 5 represents the measured flux density profiles as revealed by a calibration procedure based on photometric evaluations of the grey scales of the exposed film. The calibration procedure is described in detail in [29]. The flux density profiles of Fig.5 are not compatible with the simple Bean model which predicts linear profiles. The volume pinning forces as derived from the flux density gradients show a continuous increase towards the direction to the center and a maximum of $2.10^7$ N/m$^3$ at the boundary to the Meissnerstate.

## 6.CONCLUSIONS

The previous chapters show that quantitative high resolution MOI opens an access to a more detailed understanding of magnetic properties of superconductors. The MOI technique using

Eu-chalkogenides as active MOL's has the big advantage of high spatial resolution, the possibility of investigations of dynamical properties and quantitative analysis of flux density gradients and pinning forces as well. The disadvantage of this technique is the limitation to the temperature range < 10K and the magnetic (1 mT) and spatial resolution ( 1μm ). Nevertheless, during the period 1970– 1980 all fundamental principles mandatory for a successful experimental technique to study vortex matter has been established.. The present standard of MOI resting on the principles developed 1970-1980 appear to be a mature technology. The introduction of garnet films as magneto-optical media [31] in conjunction with modern digital image processing techniques represent a major step towards an easy-to-use technique with still room for improvements. One of the examples are the visualization of single flux-lines in $NbSe_2$ by Goa et.al. by MOI [30]. Further steps towards the goal of imaging single flux-lines in HTS materials by MOI could be expected if another class of magneto-optical layers would be available with Verdet constants exceeding those of the Eu-compounds. This would allow to combine the advantage of MOI using single magneto-optical layers for high spatial resolution with the high magnetic field resolution of the garnet films. Additionally the application of techniques based on near field optical microscopy could be a strategy to overcome the limitations of spatial resolution.

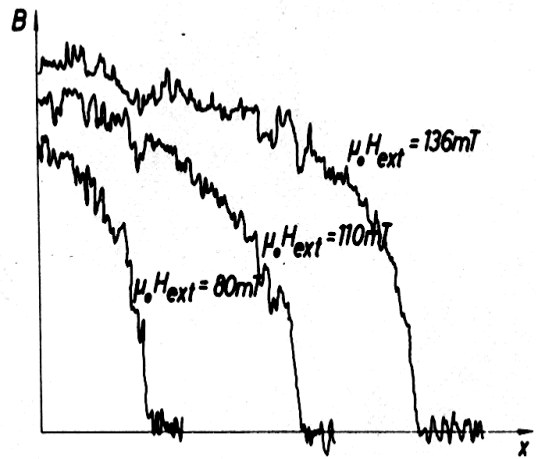

*Figure 5. Flux-density gradients of a Nb disc at different external magnetic fields at 4.5 K.*